\documentstyle[aps,preprint,epsfig]{revtex}
\begin{document}
\tightenlines
\draft

\title{An effective field theory of QCD at high density}

\author{Deog Ki Hong\footnote{
E-mail address: {\tt dkhong@hyowon.cc.pusan.ac.kr}
}}

\address{
Department of Physics, Pusan National University,
Pusan 609-735, Korea\footnote{Permanent address}
\protect\\   
Lyman Laboratory of Physics,
Harvard University, Cambridge, MA 02138
\protect\\   
Physics Department, Boston University,
Boston, MA 02215, USA}

\maketitle

\begin{abstract}
We derive a (Wilsonian) effective field theory of QCD at finite
density by
integrating out the states in the Dirac sea
when the chemical potential
$\mu\gg\Lambda_{\rm QCD}$.
The quark-gluon coupling is effectively (1+1)-dimensional and
the theory contains four-quark
operators which become relevant as we approach to the Fermi sea.
By calculating the one-loop vacuum polarization tensor
in the effective theory, we find
the electric gluons have a screening mass, $M\sim
g_s\mu$, while the static magnetic gluons are unscreened.
We then investigate the gap equations for color anti-triplet Cooper
pairs by including both gluon-exchange interactions
and the marginal four-quark interactions in the effective theory.
\end{abstract}



Quantum chromodynamics (QCD) has become a undisputed
theory for the strong interaction of quarks and gluons. However,
its direct experimental evidences are still limited to the high energy
region where quarks and gluons are weakly coupled.
At low energy where they couple
strongly,  we have only indirect evidences
due to the calculational incapability that QCD describes the
strong interaction.

Recently, QCD at high density has
been studied intensively, since it not only offers a nonperturbative
test but also exhibits rich novel phases which might be
accessible in the core of neutron stars or in a quark-gluon plasma
in heavy ion
collisions~\cite{bailin,shuryak,wilczek,wilczek2,nick,rho}.
It has
been argued that at a sufficiently high density
the color symmetry is spontaneously broken,
leading to color superconductivity.

In this letter, we derive a (Wilsonian) effective theory of QCD
at high density in loop expansion,
after integrating out the irrelevant degrees of
freedom at low energy. The effective theory includes
new quark-gluon couplings generated by the exchange of states
in the Dirac sea, which induce marginal four-fermion operators
at matching one-loop amplitudes, as in the effective theory of
QED under external magnetic fields~\cite{my}.
The marginal four-quark operators become relevant as we scale
down to the Fermi sea.
In the effective theory, the gluon-quark couplings are
effectively (1+1)-dimensional and the electric gluons, $A_0^a(x)$,
are screened in the static limit. At energy below the screening mass,
the only relevant interactions for quarks are the four-quark operators
with opposite momenta and the interactions with magnetic gluons.
We then solve the gap equations for Cooper pairs in the color
anti-triplet channel by long-range color magnetic
interactions, screened color electric interactions,
and the marginal four-quark interactions.


A system of degenerate quarks with a fixed baryon number is described
by the QCD Lagrangian density with a chemical potential $\mu$,
\begin{equation}
{\cal L}_{\rm QCD}=\bar\psi
i\!\not\!D\psi-{1\over4}F_{\mu\nu}^aF^{a\mu\nu}+\mu
\bar\psi\gamma_0\psi,
\label{lag}
\end{equation}
where the covariant derivative
$D_{\mu}=\partial_{\mu}+ig_sA_{\mu}^aT^a$ and
we neglect the mass of quarks for simplicity.
(Later, we will consider the effect of quark mass.)

For a sufficiently high density such that $\mu\gg\Lambda_{\rm QCD}$,
one can derive systematically a (Wilsonian)
effective theory of QCD, where $\Lambda_{\rm QCD}/\mu$ provides a
useful expansion parameter.
At an energy scale just below the chemical potential
$\mu$ $(\gg\Lambda_{\rm QCD})$, quarks are
almost free and their energy spectrum is  given  by an eigenvalue
equation,
\begin{equation}
\left(\vec\alpha\cdot \vec p-\mu\right)\psi_{\pm}=E_{\pm}\psi_{\pm},
\end{equation}
where $\vec\alpha=\gamma_0\vec\gamma$ and $\psi_{\pm}$ denote the
energy eigenfunctions with eigenvalues $E_{\pm}=-\mu\pm \left|\vec
p\right|$, respectively. At low energy $E<\mu$, the states $\psi_+$
near the Fermi surface,  $|\vec p|\sim\mu$, are easily excited
but $\psi_-$, which correspond to the states in the Dirac sea,
are completely decoupled due to the
presence of the energy gap $\mu$ provided by the Fermi sea.
Therefore the right degrees of freedom below $\mu$ consist of
gluons and $\psi_+$ only.

Since at very low energy ($E\ll\mu$) quarks are almost
on-shell and move at the Fermi momentum ${\vec p}_F=\mu{\vec v_F}$, we
may decompose the momentum of quarks into the Fermi momentum and a
residual momentum as
\begin{equation}
p_{\mu}=\mu v_{\mu}+l_{\mu},
\end{equation}
where the residual momentum $|l_{\mu}|<\mu$ and $v^{\mu}=(0,\vec
v_F)$.
Now, we decompose the quark fields as
\begin{equation}
\psi(x)=\sum_{\vec v_F} \left[e^{i\mu\vec v_F\cdot \vec x}
\psi_+(\vec v_F,x)+e^{i\mu\vec v_F\cdot \vec x}\psi_-(\vec v_F,x)\right],
\end{equation}
where $\vec\alpha\cdot\vec v_F\psi_{\pm}(\vec v_F,x)=
\pm \psi_{\pm}(\vec v_F,x)$.
The quark Lagrangian in Eq.~(\ref{lag}) then becomes
\begin{eqnarray}
\bar\psi \left(i\!\not\!D+\mu\gamma^0\right)\psi&=&\sum_{\vec
v_F}\left[\bar\psi_+(\vec
v_F,x)i\gamma^{\mu}_{\parallel}D_{\mu}\psi_+(\vec v_F,x)
+\bar\psi_-(\vec
v_F,x)\gamma^{0}\left(2\mu+iD_{\parallel}\right)\psi_-(\vec
v_F,x)\right]\nonumber\\
&+&\sum_{\vec v_F}\left[\bar\psi_-(\vec
v_F,x)i\!\not\!D_{\perp}\psi_+(\vec v_F,x)+{\rm h.c.}\right]
\end{eqnarray}
where $\gamma^{\mu}_{\parallel}\equiv(\gamma^0,\vec v_F\vec
v_F\cdot\vec \gamma)$, $\gamma^{\mu}_{\perp}=\gamma^{\mu}-
\gamma^{\mu}_{\parallel}$, $D_{\parallel}=\bar V^{\mu}D_{\mu}$ with
$\bar V^{\mu}=(1,-\vec v_F)$, and
$\not\!D_{\perp}=\gamma^{\mu}_{\perp}D_{\mu}$.

At low energy, since the fast modes $\psi_-$ are decoupled,
we integrate out
all the fast modes $\psi_-$ to derive the low energy effective
Lagrangian by matching all the one-light particle irreducible
amplitudes containing gluons and $\psi_+$ in loop expansion.
The effects of fast modes will appear in
the quantum corrections to the couplings of low energy interactions.
At tree-level, the matching is equivalent to eliminating $\psi_-$
in terms of equations of motion;
\begin{equation}
\psi_-(\vec v_F,x)=-{i\gamma^0\over 2\mu
+iD_{\parallel}}\not\!D_{\perp}\psi_+(\vec v_F,x)=-
{i\gamma^0\over 2\mu}\sum_{n=0}^{\infty}\left(-{iD_{\parallel}\over
2\mu}\right)^n\not\!D_{\perp}\psi_+(\vec v_F,x).
\label{eom}
\end{equation}
Therefore, the tree-level Lagrangian for $\psi_+$ becomes,
using $(1-\vec\alpha\cdot\vec v_F)/2\gamma^{\mu}
(1+\vec \alpha\cdot\vec v_F)/2=\gamma^0V^{\mu}
(1+\vec \alpha\cdot\vec v_F)/2$
with $V^{\mu}=(1,\vec v_F)$,
\begin{equation}
{\cal L}_{\rm eff}^0=\sum_{\vec v_F}\left[
\psi_+^{\dagger}V^{\mu}D_{\mu}\psi_+-{1\over2\mu}\psi_+^{\dagger}
(\not\!D_{\perp})^2\psi_++\cdots\right],
\end{equation}
where the ellipsis denotes terms with higher derivatives.
We see that in the high density limit the quark-gluon couplings are
effectively (1+1)-dimensional: In the leading order, to quarks of
Fermi velocity $\vec v_F$ only the $V\cdot A$ component of gluons
couple and $A^{\mu}_{\perp}=(0,\vec A)$ with $\vec v_F\cdot\vec A=0$
does not couple to the color charge of $\psi_+$ but
to its color magnetic moment, which is suppressed by $1/\mu$.


By the tree-level matching shown in Fig.~1,
we obtain a vertex of two gluons and two quarks, generated by the
exchange of $\psi_-$,
\begin{equation}
{\cal L}^{\rm eff}_{int}=-{g_s^2\over2\mu}\sum_{\vec
v_F}\bar\psi_+\!\not\! A_{\perp}\gamma_0\!\not\!
A_{\perp}\psi_++\cdots,
\end{equation}
where the ellipses
denote terms containing more powers of gluons and derivatives. (From
now on $\psi$ will denote the slow modes $\psi_+$.)

Now, consider the one-loop matching of a four-quark amplitude.
(See Fig.~2.)
In QCD, the amplitude is ultraviolet (UV) finite but
infrared (IR) divergent, while it
is both UV and IR divergent in the effective theory. Since the IR
divergence is same in both theories, we need a
UV counter term in the effective theory to match the amplitude,
which is a four-quark operator.  The one-loop four-quark
amplitude in the effective theory is
\begin{eqnarray}
A_4^{\rm eff} &=& {i\sqrt{2} g_s^4\over 128\pi^2\mu^2}
(2\pi)^4\delta^4(p_3+p_4-p_1-p_2) \left[{1\over
\epsilon}-\gamma+\ln
4\pi-2-\ln\left({-l^2\over\Lambda^2}\right)\right]  \nonumber\\ &
& \times
\left[u^{\dagger}_t(p_3)u_s(p_1)u^{\dagger}_v(p_4)u_u(p_2) \cdot
\left({13\over9}\delta^S_{us;tv}-{2\over9}\delta^A_{us;tv}\right)
\right.
\\
& & \left.
+u^{\dagger}_t(p_3)\gamma_5u_s(p_1)u^{\dagger}_v(p_4)\gamma_5u_u(p_2)
\cdot \left(\delta^S_{us;tv}-2\delta^A_{uv;ts}\right)\right] +{\rm
c.t.} \nonumber
\end{eqnarray}
where $l$ is the external momentum transfer, $\Lambda$ is a
renormalization point, and c.t. denotes the counter terms.
$u,s,t,v$ are color indices and
$\delta^S_{us;tv}=(\delta_{uv}\delta_{ts}+\delta_{ut}\delta_{sv})
/\sqrt{2}$,
$\delta^A_{us;tv}=(\delta_{uv}\delta_{ts}-\delta_{ut}\delta_{sv})
/\sqrt{2}$.
We find that the new quark-gluon coupling generates
effective four-quark operators at one-loop;
\begin{eqnarray}
S_{\rm eff}&\ni& {1\over
2\mu^2}
\int \prod_{i=1}^4{{\rm d}^4p_i\over (2\pi)^4}(2\pi)^4\delta^4
(p_3+p_4-p_1-p_2)
\left[
\psi^{\dagger}_t(p_3)\psi_s(p_1)\psi_v^{\dagger}(p_4)
\psi_u(p_2)
\left(g_{\bar 3}\delta^A_{us;tv}-g_6\delta^S_{us;tv}\right)
\right.\nonumber\\
& &\left.
+\psi^{\dagger}_t(p_3)\gamma_5\psi_s(p_1)\psi_v^{\dagger}
(p_4)\gamma_5\psi_u(p_2)
\left(h_{\bar 3}\delta^A_{us;tv}-h_6\delta^S_{us;tv}\right)
\right].
\end{eqnarray}
At the matching scale $\Lambda=\mu$, the couplings are of order of
$\alpha_s^2$. As in the BCS
superconductivity~\cite{pol}, this four-quark operator is marginal
under the scale transformation toward the Fermi surface,
$l_{\parallel}\to s l_{\parallel}$ with $s<1$, only when
the incoming quarks have opposite Fermi momenta,
$p_1+p_2=0+O(l)$, where we write the quark momenta
$p_i^{\mu}=\mu v_i^{\mu}+l_i^{\mu}$ as before.
Therefore, if this four-quark operator leads to a condensate, it must
be in the s-channel, $\left<\psi_s(p_1)\psi_u(p_2)\right>\ne0$ or
$\left<\psi_s(\vec v_F,x)\psi_u(-\vec v_F,x)\right>\ne0$.
Note also
that we do not have marginal operators containing $\bar
\psi_t(\vec v_F,x)\psi_s(\vec v_F,x)$ because it vanishes identically
for the slow modes $\psi_+$.  Therefore, one can easily understand in
the effective theory that
the chiral symmetry will be restored
as the chemical potential becomes large, if any
translationally invariant order parameter for chiral symmetry breaking
should contain the fast modes $\psi_-$~\cite{locking}.

Next, we match the gluon two-point amplitude at one-loop. Since
the quarks in the Dirac sea are integrated out, we need to match
only the quark-loop correction to the amplitude. The quark-loop
contribution to the gluon two-point function in QCD at finite
density and at zero temperature is calculated by
Manuel~\cite{cristiana}. It consists of two parts; one due to the
matter and the other due to the vacuum. In terms of the vacuum
polarization tensor, $\Pi^{\mu\nu}_{ab\rm full}(p)
=\Pi{\mu\nu}_{ab\rm mat}(p)+\Pi^{\mu\nu}_{ab\rm vac}(p) $, where
$\Pi^{\mu\nu}_{ab\rm vac}$ is the quark-loop contribution when
there is no matter, $\mu=0$. For $\left|p^{\mu}\right|\ll\mu$, the
matter part of the vacuum polarization becomes for $N_f$ light
quarks,  with $M^2=N_f g_s^2\mu^2/(2\pi^2)$,
\begin{equation}
\Pi^{\mu\nu}_{ab\rm mat}
=-{iM^2\over2}\delta_{ab} \int{{\rm d}\Omega_{\vec v_F}\over 4\pi}
\left({-2\vec p\cdot\vec v_FV^{\mu}V^{\nu}\over p\cdot V
+i\epsilon \vec p\cdot\vec v_F}+g^{\mu\nu}-{
V^{\mu}{\bar V}^{\nu}+{\bar V}^{\mu}V^{\nu}\over2}\right)
\end{equation}
which is transversal,
$p_{\mu}\Pi^{\mu\nu}_{ab\rm mat}(p)=0$~\cite{cristiana}.

On the other hand,  in the effective theory the quark-loop correction
to the vacuum polarization is given as:
\begin{eqnarray}
\Pi^{\mu\nu}_{ab}(p)
&=&-g_s^2\delta_{ab} \sum_{\vec v_F}
\int {{\rm d}^4q\over (2\pi)^4}{\rm Tr}
\left[\left({1-\vec\alpha\cdot\vec
v_F\over2}\right)\gamma^{\nu}_{\parallel}
{i\over\not q_{\parallel}+i\epsilon}
\gamma^{\mu}_{\parallel}{i\over \not
q_{\parallel}+ \not p_{\parallel}+i\epsilon}\right] \nonumber \\
&=&-{i\over 8\pi}\delta_{ab}
\sum_{\vec v_F}V^{\mu}V^{\nu}
\left(1-
{p_0+\vec v_F\cdot\vec p\over
p_0-\vec v_F\cdot \vec p+i\epsilon p_0}\right)M^2.
\label{vpol}
\end{eqnarray}
To match the gluon two-point amplitudes in both theories,
we therefore need to add a term in the one-loop
effective Lagrangian,
\begin{equation}
{\cal L}_{\rm eff}\ni -{M^2\over 16\pi}\sum_{\vec v_F}A_{\perp}^{a\mu}
A_{\perp\mu}^a,
\label{add}
\end{equation}
which also ensures the gauge invariance of the effective Lagrangian
at one-loop.
Since in the effective theory
$\Pi^{00}_{ab}\simeq -iM^2\delta_{ab}$, as $p_0\to0$,
we see that the electric gluons, $A_0^a$, have a screening mass, but
the static magnetic gluons are not screened at one-loop
due to the added term Eq.~(\ref{add}),
which holds at all orders in perturbation as in
the finite temperature~\cite{kapusta,pisarski}.

Finally, we match the quark two-point amplitudes
to get a one-loop low energy (Wilsonian) effective Lagrangian
density
\begin{eqnarray}
{\cal L}_{\rm eff}&=&-{1\over4}(1+a_1)
\left(F_{\mu\nu}^a\right)^2 -
{M^2\over 16\pi}A_{\perp}^{a\mu}A_{\perp\mu}^a
+(1+b_1)\bar\psi i\gamma_{\parallel}^{\mu}D_{\mu}\psi
-{1\over2\mu}(1+c_1)
\psi^{\dagger}\left(\gamma_{\perp}\cdot D\right)^2 \psi
\nonumber \\
&+&
{1\over2\mu^2}
\left[
\left(g_{\bar 3}\delta^A_{us;tv}-g_6\delta^S_{us;tv}\right)
\psi^{\dagger}_t(\vec v_F,x)\psi_s(\vec v_F,x)
\psi^{\dagger}_v(-\vec v_F,x)\psi_u(-\vec v_F,x)
\right. \label{eff}
\\
& &\left.
+\left(h_{\bar 3}\delta^A_{us;tv}-h_6\delta^S_{us;tv}\right)
\psi^{\dagger}_t(\vec v_F,x)\gamma_5\psi_s(\vec v_F,x)
\psi^{\dagger}_v(-\vec v_F,x)
\gamma_5\psi_u(-\vec v_F,x)\right]+\cdots,
\nonumber
\end{eqnarray}
where the summation over $\vec v_F$ is suppressed
and the coefficients
$a_1,b_1,c_1$ are dimensionless and of order
$\alpha_s(\mu)$. The ellipsis denotes the irrelevant
four-quark operators and terms with more external fields and
derivatives.

As we scale further down, the effective four-quark
operators will evolve together with other operators, which can be seen
by further integrating out the high frequency modes,
$s\mu<|l_{\mu}|<\mu$. The scale dependence of the four-quark operators
has three pieces.
One is from the one-loop matching condition for the
four-quark amplitudes and the other two are from the loop corrections
to the four-quark operators, shown in Fig.~3.
Putting all contribution together, we find the one-loop
renormalization group equations for the four-quark operators to be
\begin{equation}
s{\partial \over\partial s}{\bar g}_i=-
\gamma_i\alpha_s^2-{1\over 4\pi^2}{\bar g}_i^2-
{\ln2\over12\pi}\delta_i{\bar g}_i\alpha_s,
\label{rge}
\end{equation}
where $i=(6,{\bar 3})$, $\bar g_6=-g_6$, $\bar
g_{\bar3}=g_{\bar3}$ and $\gamma_i=(\sqrt{2}/9)(13/4,1/2)$ and
$\delta_i=(-1,2)$. Since in the high density limit the
quark-gluon coupling is (1+1)-dimensional, the quarks do not
contribute to the running of the strong coupling. The one-loop
$\beta$ function for the strong coupling constant at high density
is $\beta(\alpha_s)=-11/(2\pi)\alpha_s^2$. Integrating the RG
equation (\ref{rge}), we find for $\mu\gg\Lambda$
\begin{equation}
{\bar g}_{i}(\Lambda)\simeq{2\pi\over
11}\alpha_s(\Lambda)\left[\gamma_i
- {(\ln2)^2\over144}\delta_i^2\right].
\end{equation}
At a scale much less than the chemical potential, $\Lambda\ll\mu$,
$g_{\bar3}(\Lambda)\simeq 0.04\alpha_s(\Lambda)$ and
$g_6(\Lambda)\simeq0.29\alpha_s(\Lambda)$. Similarly for $h_i$,
$\gamma_i=\sqrt{2}/2(2,1)$ and $\delta_i=(-1,2)$ and we get
$h_{\bar3}(\Lambda)\simeq 0.4\alpha_s(\Lambda)$ and
$h_6(\Lambda)\simeq 0.81\alpha_s(\Lambda)$.


At a scale below the screening mass, we further integrate out
the electric gluons, which will
generate four-quark interactions.
For quarks moving with opposite Fermi momenta, the electric-gluon
exchange four-quark interaction is given as
\begin{equation}
{\cal L}_{1g}\ni -{g_s^2(M)\over 2M^2}\sum_{\vec v_F}
\bar\psi\gamma^{0}T_a\psi(\vec v_F,x)
\bar\psi\gamma_0T_a\psi(-\vec v_F,x).
\end{equation}
Using  $T^a_{tu}T^a_{vs}
=1/2\delta_{ts}\delta_{uv}-1/6\delta_{tu}\delta_{vs}$, we find
that the four-quark couplings are shifted as $g_{\bar3}(M)\to
g_{\bar3}(M)+2\sqrt{2}g_s^2(M)/3\simeq 0.95g_s^2(M)$ and
$g_6(M)\to g_6(M)+\sqrt{2}g_s^2(M)/3\simeq0.49g_s^2(M)$, while
$h_i$'s are unchanged.

As we approach further to the Fermi surface, closer than the
screening mass $M$, the four-quark operators in the color anti-triplet
channel become stronger, because the $\beta$ function for the
attractive  four-quark operators is negative,
$\beta (g_{\bar 3})=-g_{\bar3}^2/(4\pi^2)$.
If the four-quark interaction is dominant at low energy, it leads to
vacuum instability in the infrared region  by forming a color
anti-triplet condensate or Cooper pair,
$\epsilon^{tsu}\left<\psi^{\alpha i}_s(\vec v_F,x) \psi^{\beta
j}_u(-\vec v_F,x)\right>\ne0$, where flavor indices $\alpha,\beta$ and
Dirac indices $i,j$ are  restored. The size of the condensate
driven by the four-quark interaction is
determined  by the RG invariant scale,
\begin{equation}
\Delta\simeq M\exp\left(-{4\pi^2\over 0.95g_s^2(M)}\right).
\label{rggap}
\end{equation}
But, since the long-range color-magnetic interactions
become also strong at low energy, we need to consider both
interactions to determine the Cooper-pair gap.

To describe the Cooper-pair gap equation, we introduce a charge
conjugate field,
\begin{equation}
\left(\psi_c\right)_{\alpha}=C_{\alpha\beta}\bar\psi_{\beta}(x),
\end{equation}
where $\alpha$ and $\beta$ are Dirac indices and the matrix $C$
satisfies $C^{-1}\gamma_{\mu}C=-\gamma_{\mu}^T$.
Then, we may write the inverse propagator for $\Psi(\vec
v_F,x)=(\psi(\vec v_F,x),\psi_c(-\vec v_F,x))^T$ as
\begin{equation}
S^{-1}(\vec v_F,l)= \gamma^0\pmatrix{l\cdot V &
       \Delta(l_{\parallel}) \cr
\Delta^{\dagger}(l_{\parallel})& l\cdot\bar V\cr}.
\end{equation}
The gap equation for $\Delta$ in the color-antitriplet channel
is then given in the hard dense loop (HDL) approximation as
(see Fig.~4)
\begin{equation}
\Delta(p_{\parallel})=(-ig_s)^2\int {{\rm d}^4l\over (2\pi)^4}
D_{\mu\nu}(p-l)V^{\mu}{T^a\Delta(l_{\parallel})(T^a)^T\over l_{\parallel}^2
-\Delta^2(l_{\parallel})}{\bar V}^{\nu}
+i{g_{\bar3}\over \mu^2}\int {{\rm d}^4l\over (2\pi)^4}
{\Delta(l_{\parallel})\over l_{\parallel}^2-\Delta^2(l_{\parallel})},
\end{equation}
where $D_{\mu\nu}$ is the gluon propagator.

The gluon propagator is given in the HDL approximation
as,\footnote{In the Schwinger-Dyson equation the loop
momentum should take the whole range up to the ultraviolet cutoff,
which is the chemical potential $\mu$ in the case of high density effective
theory. Hence the gluon propagator includes both magnetic and electric
gluons.} following the notations used by Sch\"afer and Wilczek~\cite{sw99},
\begin{equation}
iD_{\mu\nu}(k)={P_{\mu\nu}^{T}\over k^2-G}+{P^L_{\mu\nu}\over k^2-F}
-\xi {k_{\mu}k_{\nu}\over k^4},
\end{equation}
where $\xi$ is the gauge parameter and
the projectors are defined by
\begin{eqnarray}
P^T_{ij}&=&\delta_{ij}-{k_ik_j\over |\vec k|^2},
\quad P_{00}^T=0=P_{0i}^T\\
P^L_{\mu\nu}&=&-g_{\mu\nu}+{k_{\mu}k_{\nu}\over k^2}-P^T_{\mu\nu}.
\end{eqnarray}

In the weak coupling limit, $|k_0|\ll|\vec k|$ and thus
\begin{eqnarray}
F(k_0,\vec k)\simeq M^2,\quad
G(k_0,\vec k)\simeq {\pi\over 4}M^2{k_0\over |\vec k|}.
\end{eqnarray}
Since the gap has to be fully antisymmetric in color indices, we get
\begin{eqnarray}
T_{tu}^a\Delta_{uv}(T^a)^T_{vs}=
\left({1\over 2}\delta_{tv}\delta_{us}-{1\over6}\delta_{tu}\delta_{vs}
\right)\Delta_{uv}=-{2\over 3}\Delta_{ts}
\end{eqnarray}
After Wick-rotating into Euclidean space, the gap equation becomes
\begin{eqnarray}
\Delta(p_{\parallel})\!&=&\!\int{d^4q\over (2\pi)^4}
\left[-{2\over3}g_s^2
\left\{{V\cdot P^T\cdot \bar V
\over (p-q)_{\parallel}^2+
\vec q_{\perp}^2+{\pi\over4}M^2|p_0-q_0|/
|\vec p-\vec q|}\right.\right.\nonumber\\
&-&\left.\left.{1\over (p-q)_{\parallel}^2+{\vec q_{\perp}}^2+M^2}
-\xi{(p-q)_{\parallel}^2\over
(p-q)^4}\right\}+{g_{\bar3}\over \mu^2}\right]
{\Delta(q_{\parallel})\over q_{\parallel}^2+\Delta^2(q_{\parallel})}.
\label{gap-hdl}
\end{eqnarray}
Note that the main contribution~\footnote{
The hard dense loop approximation is therefore consistent since
$\Delta\ll M$. I thank Steve Hsu for explaining this point.}
to the integration comes from
the loop momenta in the region $q_{\parallel}^2\sim \Delta^2$ and
$|\vec q_{\perp}|\sim M^{2/3}\Delta^{1/3}$. Therefore, we find that
the leading contribution is by the first term due to the Landau-damped
magnetic gluons.
For this momentum range, we can take
$|\vec p-\vec q|\sim |\vec q_{\perp}|$ and
\begin{eqnarray}
V\cdot P^T\cdot{\bar V}=
-v_F^iv_F^j\left(\delta_{ij}-{\hat k}_i{\hat k}_j\right)
=-1+O\left({\Delta^{4/3}\over M^{4/3}}\right).
\end{eqnarray}
We also note that the term due to the four-Fermi operator is negligible,
since $g_{\bar3}\sim g_s^4$ at the matching scale $\mu$.

Neglecting $(p-q)_{\parallel}^2$
in the denominator, the gap equation becomes at the leading order
in the weak coupling expansion and $1/\mu$ expansion
\begin{eqnarray}
\Delta(p_{\parallel})={2g_s^2\over3}\int{d^4q\over (2\pi)^4}
\left[
{1\over \vec q_{\perp}^2+{\pi\over4}M^2|p_0-q_0|/|\vec q_{\perp}|}
+{1\over {\vec q_{\perp}}^2+M^2}
+\xi{(p-q)^2_{\parallel}\over |\vec q_{\perp}|^4}\right]
{\Delta(q_{\parallel})\over q_{\parallel}^2+\Delta^2(q_{\parallel})}.
\end{eqnarray}
The $\vec q_{\perp}$ integration can now be performed easily to get
\begin{eqnarray}
\Delta(p_{\parallel})={g_s^2\over 9\pi}
\int{d^2q_{\parallel}\over (2\pi)^2}{\Delta(q_{\parallel})\over
q_{\parallel}^2+\Delta^2}\left[\ln\left(
{\mu^3\over {\pi\over4}M^2|p_0-q_0|}\right)
+{3\over2}\ln\left({\mu^2\over M^2}\right)+{3\over2}\xi\right].
\end{eqnarray}
We see that in this approximation $\Delta(p_{\parallel})\simeq
\Delta(p_0)$. Then, we can integrate over $\vec v_F\cdot\vec q$
to get
\begin{eqnarray}
\Delta(p_0)&=& {g_s^2\over 36\pi^2}\int_{-\mu}^{\mu}dq_0
{\Delta(q_0)\over \sqrt{q_0^2+\Delta^2}}
\ln\left({{\bar\Lambda}\over |p_0-q_0|}
\right)
\label{gapf}
\end{eqnarray}
where $\bar\Lambda=4\mu/\pi\cdot (\mu/M)^5e^{3/2\xi}$.
If we take $\Delta\simeq \Delta(0)$ for a rough estimate of the gap,
\begin{equation}
1={g_s^2\over 36\pi^2}
\left[\ln\left({\bar\Lambda\over\Delta}\right)\right]^2
\quad
{\rm or}\quad
\Delta\simeq\bar\Lambda \exp\left(-{6\pi\over g_s}\right).
\end{equation}
As was done by Son~\cite{son}, one can convert the
Schwinger-Dyson gap equation (\ref{gapf})
into a differential equation to take into
account the energy dependence of the gap.
Approximating the logarithm in the gap equation as
\begin{equation}
\ln\left|p_0-q_0\right|\simeq
\left\{
\begin{array}{ll}
\ln\left|p_0\right| & \mbox{if $\left|p_0\right|>\left|q_0\right|$},\\
\ln\left|q_0\right| & \mbox{otherwise},
\end{array}
\right.
\end{equation}
we get
\begin{equation}
p\Delta^{\prime\prime}(p)+\Delta^{\prime}(p)+{2\alpha_s\over9\pi}
{\Delta(p)\over \sqrt{p^2+\Delta^2}}=0,
\label{diff}
\end{equation}
where $p\equiv p_0$. The solution to the differential equation
(\ref{diff}) is found to be at $p=\Delta(p)$
\begin{equation}
\Delta\simeq \bar\Lambda\exp\left(-{\pi\over\nu}+1+O(\nu^2)\right),
\end{equation}
where $\nu=\sqrt{8\alpha_s/(9\pi)}$. The gap is therefore given as
at the leading order in the weak coupling expansion\footnote{
The gauge-parameter dependent term is subleading
in the gap equation (\ref{gap-hdl}).
Since the gap has to be gauge-independent, the gauge parameter
dependence in the prefactor will disappear if one includes
the higher order corrections.}
\begin{equation}
\Delta=c\cdot{\mu\over g_s^5}\exp\left(-{3\pi^2\over\sqrt{2}g_s}\right),
\end{equation}
where $c=2^7\pi^4N_f^{-5/2}e^{3/2\xi+1}$. This agrees with
the RG analysis done by Son~\cite{son} (see also~\cite{hsu99})
and also with
the Schwinger-Dyson approach in full QCD~\cite{hmsw,sw99,pr99}.
The $1/g_s$ behavior of the exponent of the gap at high density is due to
the double logarithmic divergence in the gap equation~(\ref{gap-hdl}),
similarly to the case of chiral symmetry breaking under external magnetic
fields~\cite{miransky,hong3}.
In addition to the usual logarithmic divergence
in the quark propagator as in the BCS superconductivity,
there is another logarithmic divergence due to the long-range gluon
exchange interaction, which occurs when the gluon loop momentum is colinear
to the incoming quark momentum ($\vec q_{\perp}\to0$).

The color
anti-triplet condensate has an interesting flavor and Lorentz
structure as discussed in~\cite{shuryak,wilczek,rho,pisarski}. So far
we have neglected quark masses, since they are irrelevant at  scales
higher than the quark masses, namely at $\Lambda>m_s$, the mass of
strange quark in the three flavor case. But, if  $g_{\bar3}$ or
$g_s$ do not become strong enough to form a condensate  at the scale
$\Lambda=m_s$, the strange quark becomes irrelevant in the  low energy
effective theory and decouples from the condensate.
Furthermore, if the condensate does not form at low energy scale where
the instanton effects are important, one may expect the
instanton contribution to the condensate will be quite
significant~\cite{shuryak,wilczek}. But, as analyzed in~\cite{nick} in
detail, the instanton generated operators are not dominant when the
chemical potential is quite large.

In conclusion, we have derived a (Wilsonian) low energy
effective theory of QCD at high density
in loop expansion by integrating out the
antiparticles. The effective theory contains marginal four-quark
operators which become relevant as we approach to the Fermi sea.
We also calculate the screening mass for gluons and find that
electric gluons have screening mass, $M\sim g_s\mu$, while the
magnetic gluons are not screened in the static limit.
Finally,  in the effective theory, we solve the gap equation for
the Cooper pairs in the color anti-triplet channel in the hard dense
loop approximation. We find that the gap is given
as $\Delta\sim \mu g_s^{-5}\exp[-3\pi^2/(\sqrt{2}g_s)]$
up to a numerical constant in the prefactor, verifying recent results.

\acknowledgments

The author is grateful to Sekhar Chivukula, Andy Cohen, Nick Evans,
Steve Hsu, Roman Jackiw, Rob Pisarski, Dirk Rischke, and
Raman Sundrum for useful discussions.
This work was supported in part by the KOSEF through SRC program of
SNU-CTP and also by the academic research fund of Ministry of
Education, Republic of Korea, Project No. BSRI-98-2413.

\newpage
\begin{figure}
\vskip 0.2in
\centerline{\epsfbox{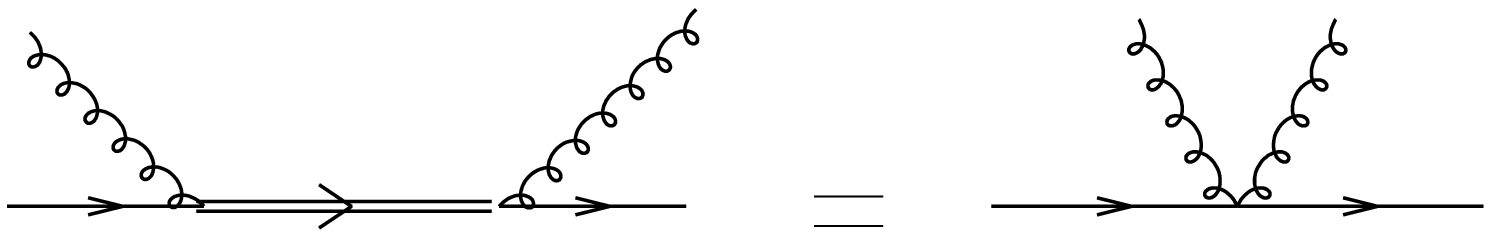}}
\vskip 0.2in
\caption{The tree-level matching condition. Wiggly lines denote gluons;
solid lines, states near the Fermi surface;
and double solid line, states in the Dirac sea. }
\end{figure}
\vskip 0.2in
\begin{figure}
\centerline{\epsfbox{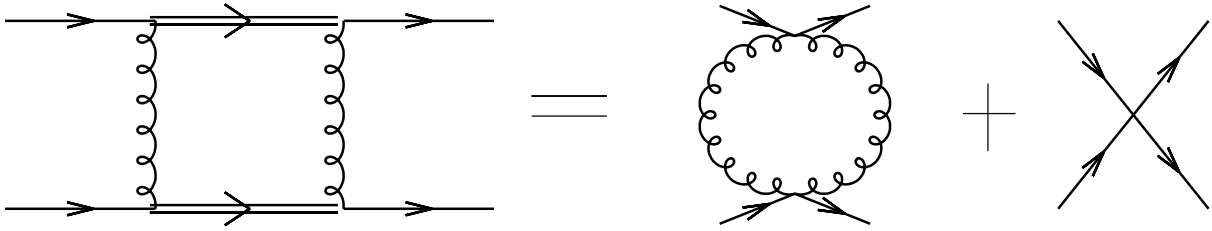}}
\vskip 0.2in
\caption{The one-loop matching condition for a four-quark amplitude.}
\end{figure}
\vskip 0.2in
\vfill
\eject
\begin{figure}
\centerline{\epsfbox{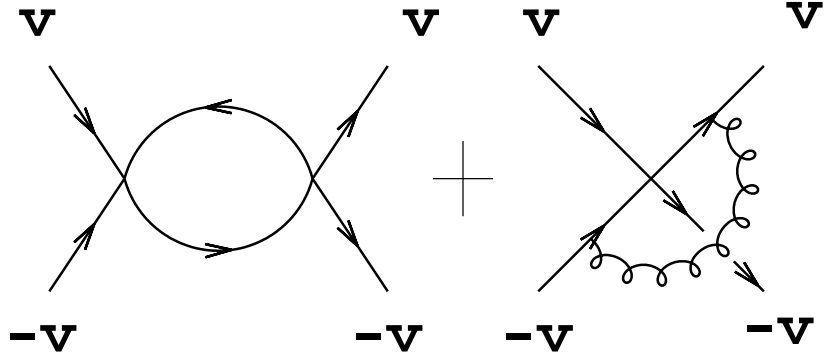}}
\vskip 0.2in
\caption{One-loop corrections to four-quark operators}
\end{figure}
\vskip 0.2in
\begin{figure}
\centerline{\epsfbox{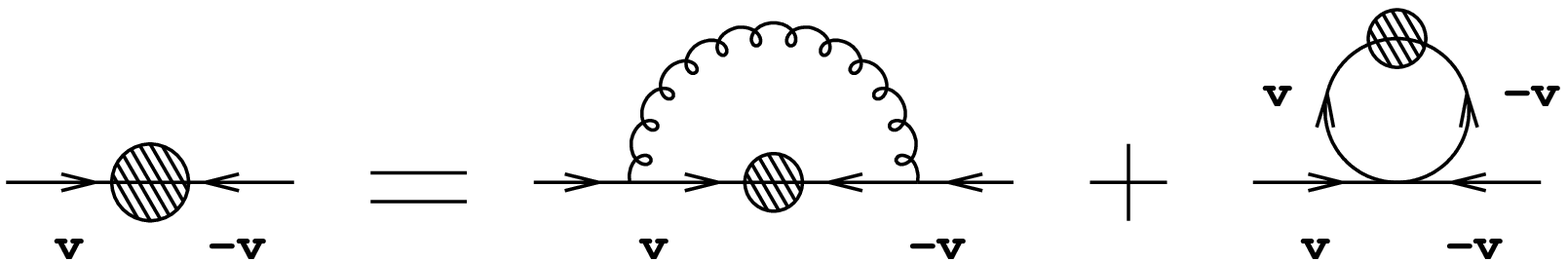}}
\vskip 0.2in
\caption{The gap equation for the color anti-triplet Cooper pairs.}
\end{figure}


\begin{references}
\bibitem{locking}An interesting possibility for chiral symmetry
breaking at high density by color-flavor locking for the three
light flavors  is suggested in~\cite{wilczek2}.
\end{references}
\end{document}